\documentclass[reprint,prd,nofootinbib]{revtex4-1}
\usepackage{graphicx}
\usepackage{xcolor}
\usepackage{amsmath}
\usepackage[abbreviations]{siunitx}
\usepackage{subfigure}
\usepackage[hidelinks]{hyperref}
\usepackage{multirow}
\def\equationautorefname#1#2\null{%
  Eq.\;(#2\null)%
}
\def\figureautorefname#1\null{%
  Fig.#1\null
}

\begin{document}

\title{Electrically charged supermassive twin stars}
\author{Victor P. {\sc Gon\c{c}alves}}
\affiliation{High and Medium Energy Group, Instituto de F\'{\i}sica e Matem\'atica,  Universidade Federal de Pelotas (UFPel)\\
Caixa Postal 354,  96010-900, Pelotas, RS, Brazil}

\author{Jos\'e C. {\sc Jim{\'e}nez}}
\affiliation{Instituto de F\'\i sica, Universidade de S\~ao Paulo (USP),\\
Rua do Mat\~ao, 1371, Butant\~a, 05508-090, SP, Brazil}

\author{Lucas {\sc Lazzari}}
\affiliation{High and Medium Energy Group, Instituto de F\'{\i}sica e Matem\'atica,  Universidade Federal de Pelotas (UFPel)\\
Caixa Postal 354,  96010-900, Pelotas, RS, Brazil}

\begin{abstract}

By assuming that ultra dense hybrid neutron stars are endowed with a distribution of electric charge, we study the corresponding twin star solutions and their properties resulting from a sharp first order transition from confined hadronic to a deconfined quark phase. Two distinct quark matter equations of state with increasing stiffness are considered and the values for the maximum gravitational masses of the hadronic and hybrid twin configurations are obtained for different values of the total electric charge. Interestingly, our calculations indicate that sharp transitions make charged twin hybrid stars more massive than their neutral counterparts, and that the 
\,{$\SI{2}{M_{\odot}}$ constraint from PSR J0740+6620} is surpassed for standard values of electric charge and can be considered stable only satisfying $\partial M/\partial \epsilon_0 > 0$. In particular, our charged stellar models reach masses even higher than the unknown compact object measured in the GW190814 event.
\end{abstract}



\maketitle
\section{Introduction}
\label{sec:intro}

Compact stars are an unique laboratory to study the strong interactions theory in the ultra dense regime, beyond the nuclear regime, where a phase transition between the hadronic and deconfined quarks degrees of freedom is expected to occur (for a recent review see, e.g. Ref.~\cite{Dexheimer:2020zzs}). If such a phase transition is sharp, a third family of compact stars can be present in the mass-radius diagram, with smaller radii but similar masses to those predicted for the disconnected second family branch~\cite{Gerlach:1968zz,Kampfer:1981yr, Kampfer:1981zmq, Glendenning:1998ag,Schertler:2000xq, Blaschke:2013ana, Benic:2014jia,Alvarez-Castillo:2014dva}. The description of these twin star configurations has been a theme of debate in recent years~\cite{Alford:2013aca,Alvarez-Castillo:2017qki,Christian:2017jni,Ayriyan:2017nby,Montana:2018bkb,Alvarez-Castillo:2018pve,Blaschke:2020vuy,Rather:2020lsg,Dexheimer:2020rlp,Jakobus:2020nxw,Christian:2020xwz,Tan:2021ahl}, mainly motivated by the perspective that future X-ray astronomy will be able to measure the masses and radii of compact stars with a high precision. The studies performed, e.g., in Refs.~\cite{Alvarez-Castillo:2017qki,Christian:2017jni,Christian:2020xwz,Tan:2021ahl} indicate that the properties of the twin stars is strongly sensitive to the treatment of the phase transition in ultra dense matter. As a consequence, a future discovery of twin stars will be important to improve our understanding about the equation of state (EoS) of compact objects.

One of the main constraints in the EoS comes from the mass of the most massive pulsars observed ($ \approx \SI{2}{M_{\odot}}$), since a viable EoS has to be able to generate compact stars with a maximum mass higher than this constraint. In recent years, additional constraints have been obtained by the analysis of gravitational wave data for the coalescence of compact binary systems. In particular, such data has been used to obtain information about the sources, such as the progenitor masses ($M_{\rm prog1,2}$). For instance, the gravitational wave event GW170817~\cite{ligo2017}, apart from marking the beginning of the era of multimessenger astronomy~\cite{LIGOScientific:2017ync,Bailes:2021tot}, consistently allowed to identify the progenitors~\cite{ligo2017} as being neutron stars (NS) having $M_{\rm prog, 1}\in{[1.36,1.60]}\si{M_{\odot}}$ and $M_{\rm prog, 2}\in{[1.17,1.36]}\si{M_{\odot}}$. The interpretation of these compact stars as being composed by hadronic and/or quark degrees of freedom is still a theme of debate. On the other hand,  gravitational wave events without multimessenger counterpart and having one progenitor with $2.2<M_{\rm prog,1}/\si{M_{\odot}}<5$ are considerably more difficult to interpret\footnote{Notice that the recent events GW200105 and GW200115~\cite{LIGOScientific:2021qlt} do not belong to this class since $M_{\rm prog,1}\ll{M}_{\rm prog,2}$ and are easily characterized as NS-BH mergers.}. For instance, the low-mass progenitor ($M_{\rm prog,1}\in [1.61, 2.52]\si{M_{\odot}}$) of GW190425~\cite{LIGOScientific:2020aai} is likely a NS, but not conclusively. Another example is the event GW190814~\cite{GW190814}, since its lightest progenitor is an unknown object having $M_{\rm prog,1}\in[2.50,2.67]\si{M_{\odot}}$, possibly being a very light BH or a supermassive NS. Interestingly, if assumed to be a NS, it puts severe constraints on a large set of EoSs for dense matter\footnote{Many standard pure quark and nuclear-matter EoSs are able to reach this mass limit although not satisfying gravitational wave constraint for the radius of $\SI{1.4}{M_{\odot}}$, see e.g. Refs. \cite{Huang:2020cab,Das:2020dcq}.}. Very recently, the analysis of these supermassive compact stars has been considered to constrain twin stars, with the results derived in Ref.~\cite{Christian:2020xwz} pointing out that a mass constraint of $\SI{2.5}{M_{\odot}}$ ruled out all twin star solutions, while Ref.~\cite{Tan:2021ahl} concluded that these ultra heavy NSs can be in a twin-mass configuration but the two stable branches must be connected. 

In this paper, we extend these previous studies by considering the impact of the electric charge on the structure properties of hybrid compact stars, with particular emphasis on the ultra-dense branch, i.e. charged twin stars. We stress that although our study is strongly motivated by the higher masses reached by charged stellar configurations, specially in the view of
the possibility that the unknown object in GW190814 to be a charged hybrid star, our main aim is to investigate the generic effects of electric charge on the structure and stability of hybrid stars with sharp phase transitions. For the particular case of the GW190814 unknown object, the electric charge might be acquired during the coalescence process. In fact, several authors have discussed in recent years the possibility that during the gravitational collapse or the coalescence process of a compact binary system, matter can acquire a large amount of electric charge and, consequently, the structure equations must be solved taking into account an electric charge distribution inside the star. In particular, in Ref.~\cite{lazzari2020}, two of the authors have studied the properties of charged strange stars considering an EoS based on perturbative QCD and demonstrated that the presence of an electric charge distribution implies that the maximum mass is larger in comparison to the neutral counterpart, in agreement with the results obtained in Ref.~\cite{arbanil2015} using a simplified approach (see also Refs.~\cite{brillante2014,Panotopoulos:2019wsy,Panotopoulos:2020hkb,Jasim:2021cft}). 
Electrically charged hybrid stars were previously studied in Ref.~\cite{brillante2014}. However, these  authors considered a smooth phase transition with a mixed phase, which does not generate twin stars. In contrast, in this work, we will consider the presence of an electric charge distribution within the compact hybrid star and assume that a sharp first order phase transition occurs.
 The impact of the electric charge will be investigated for two distinct phenomenological treatments of the phase transition, which give rise to twin star solutions, and the structural properties of the charged compact stars will be obtained. Besides, we will see that our findings indicate that the presence of an electric charge distribution implies that twin stars with masses larger than $\SI{2.6}{M_{\odot}}$ are stable, i.e. the associated configurations  satisfy the classic criteria $\partial M/\partial \epsilon_0 > 0$, which is the stability condition for sharp first order transitions characterized by a rapid conversion between phases ~\cite{pereira2018}. However, notice that this can only be considered an application of our generic results but not the main finding which deals with different classifications for the transitions.

This work is organized as follows. In Sec.~\ref{sec:theory}, we summarize the main aspects of the stellar structure equations for charged compact stars. Moreover, the EoSs considered in the description of the hybrid NS will be reviewed. In  Sec.~\ref{sec:res}, we present our results for the structural properties of the charged compact stars and a comparison with the neutral case will be performed. The twin star solutions will be presented for distinct values of the electric charge and different treatments of the phase transition. Finally, in Sec.~\ref{Conclusion} we will summarize our main results and conclusions.  
\section{Framework}
\label{sec:theory}
In this section, we will present a brief review of the theoretical framework necessary to investigate electrically charged compact stars (see Refs.~\cite{lazzari2020,brillante2014,arbanil2015,Panotopoulos:2020hkb,Jasim:2021cft} for extensive details). Initially, we will present the electrically-charged structure equations and the model for the electric charge distribution used in this work. In the second part, the two treatments for the phase transition will be discussed and the associated EoSs considered are presented. Finally, the predictions for the mass-radius diagram of electrically neutral stars will be shown for completeness. 

\subsection{Charged stellar structure equations}
In the last years, several authors have discussed the basic structure equations needed to derive the properties of a charged compact star~\cite{lazzari2020,brillante2014,arbanil2015,Panotopoulos:2020hkb,Jasim:2021cft}. In our analysis,  we will assume a static and spherically symmetric charged star. The stress-energy tensor will be expressed in terms of the contributions associated to the perfect fluid and to the electromagnetic field. For the case considered, this field is fully specified by the radial component of the electric field, which can be expressed in terms of the electric charge distribution. Following the approach presented in detail in, e.g. Ref.~\citep{lazzari2020}, one finds that the Einstein-Maxwell field equations imply that the charged stellar structure is determined by the following system of equations
\begin{align}
  \label{eq:TOV-q}
  \frac{\mathrm{d}q}{\mathrm{d}r} & {} = 4\pi r^2 \rho_e e^{\lambda} \,,\\
  \label{eq:TOV-m}
  \frac{\mathrm{d}m}{\mathrm{d}r} & {} = 4\pi r^2 \epsilon  + \frac{q}{r}\frac{\mathrm{d}q}{\mathrm{d}r} \,, \\
  \label{eq:TOV-p}
  \frac{\mathrm{d}p}{\mathrm{d}r} & {} = -(\epsilon + p)\left(4\pi r p + \frac{m}{r^2}-\frac{q^2}{r^3}\right)e^{2\lambda}   \nonumber \\
  & {} \quad + \frac{q}{4\pi r^4}\frac{\mathrm{d}q}{\mathrm{d}r} \,, 
\end{align}
where $\rho_e(r)$, $q(r)$ and $m(r)$ are the electric charge density, charge and mass profiles, respectively, and Eq.~(\ref{eq:TOV-p}) is the Tolman-Oppenheimer-Volkov (TOV) equation for a charged star. The metric potential $e^{-2\lambda}$ has the Reissner-Nordstr\"om form (see Refs. \cite{brillante2014,arbanil2015,lazzari2020} for more details).

The associated boundary conditions are established as follows. At the center of the star we have that the charge and mass are null $q(0)=m(0) = 0$ and we stipulate a central pressure $p(0) = p_0$. Moreover, the stellar surface ($r=R$) is defined by the point where the pressure is null, i.e., $p(R) = 0$. At this point, we have the star's total gravitational mass $m(R) = M$ and charge $q(R) = Q$, respectively. Moreover, in order to solve the structure equations, we must specify the charge distribution and the EoS that describes the matter inside the star, the latter being discussed in the next subsection. Regarding the charge distribution, as e.g. in Ref. \cite{lazzari2020}, we will assume it is proportional to the energy density, i.e., $\rho_e = \alpha\epsilon\,,$ where, in geometric units, $\alpha$ is a dimensionless proportionality constant which can be considered a free parameter in our modeling. Such distribution will be denoted $\alpha$-distribution in what follows. In this distribution, for a constant $\alpha$ one has that $Q = \alpha M$. Requiring that the compact star does not collapse into a BH, i. e. that the exterior event horizon of the Reissner-Nordstr{\"o}m spacetime has to be smaller than the stellar radius, which implies that $Q < M$, one should have that  $\alpha < 1$. As assumed in Refs.~\cite{brillante2014,lazzari2020,ray2003,ghezzi2005}, we will also consider that the micro-physical effects of the electric charge on the EoS are negligible.



\subsection{Equations of state}
As demonstrated in previous studies, as e.g. in Refs.~\cite{Alvarez-Castillo:2017qki,Christian:2017jni,Christian:2020xwz,Tan:2021ahl}, twin star configurations arise when a sharp phase transition of the QCD matter occurs. In our analysis, we will assume that hadronic matter (HM) is described by the stiffest EoS of Hebeler {\it et al.}~\cite{hebeler2013} which is obtained from effective chiral field theory, used between baryon densities of $n_{B}\in[0.6,1.1]n_{0}$, and being supplemented by the BPS EoS~\cite{baym1971} for the outer crust with $n_{B}<0.6n_{0}$. For the quark phase, i.e. $n_{B}>1.1n_{0}$, we consider two distinct approaches, which differ in the behaviour assumption for the speed of sound. In the first model, denoted CSS in what follows, we assume a  constant speed of sound. In contrast, in second one, denoted MP, we consider that the speed of sound is dependent on the density and the associated equation of state can be described using a multipolytrope approach.

\begin{figure*}[!tb]
  \centering
  \includegraphics[width = .475\textwidth]{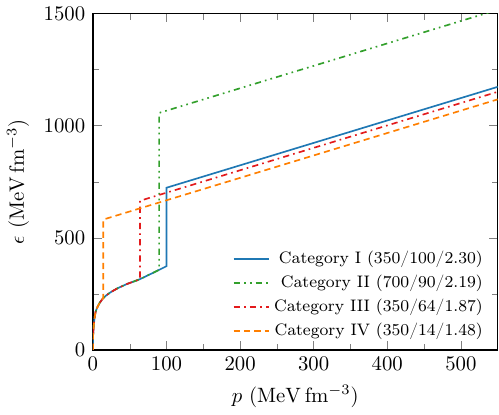}\hfill
  \includegraphics[width = .45\textwidth]{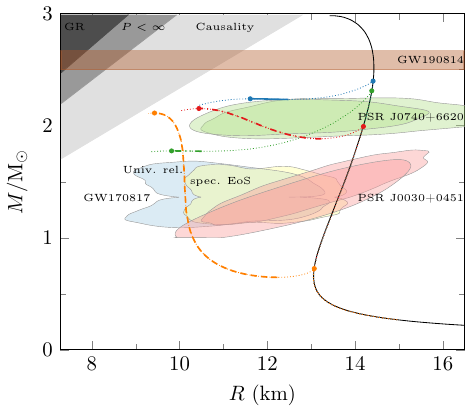}
  \caption{\label{fig:exp-MxR-4categories} {\it Left panel:} The EoS for the CSS model considering distinct values of \,{$\Delta \epsilon$, $p_t$ (both in units of $\si{MeV\,fm^{-3}}$) and the transitional baryon number density (in units of $n_0$)}, which generate twin stars in the four categories defined in Ref.~\cite{Christian:2017jni}. {\it Right panel:} Corresponding predictions for the mass-radius relation of a neutral compact star. For comparison, the pure hadronic case (black thin curve) obtained from the Hebeler {\it et al.}  stiffest EoS~\cite{hebeler2013} is presented. The maximum hadronic and hybrid masses are marked by filled circles. We also present the mass-radius constraints from the GW170817 event~\cite{ligo2017} considering Universal Relations (blue region) and a spectral EOS (yellow region which looks green when overlapped). Moreover, we add the NICER constraints~\cite{miller2019,riley2019} for PSR J0030+0451 (red regions), the updated constraint in mass~\cite{cromartie2019,fonseca2021} and radius~\cite{miller2021,riley2021} of the most massive pulsar PSR J0740+6620 (green region) 
  and the mass of the low-mass companion of GW190814~\cite{GW190814} (brown horizontal band). Finally,  the general relativity, finite pressure and causality limits~\cite{lattimer2007} are also presented in different shades of grey.}
\end{figure*}

Following Refs.~\cite{Christian:2017jni,Christian:2020xwz,Alford:2013aca}, we will consider  that the entire EoS for the CSS model is given by
\begin{equation}
  \label{eq:CSS}
  \epsilon(p) =
  \begin{cases}
    \epsilon_{\rm HM}(p) \qquad \qquad \qquad \qquad \quad ~~~~~ p < p_t\,,\\
    \epsilon_{\rm HM}(p_t) + \Delta \epsilon + c^{-2}_{\rm QM}(p-p_t) ~~~~ p > p_t\,,
  \end{cases}
\end{equation}
where  $p_{t}$ is the transitional pressure and $\epsilon_{\rm HM}(p_t)$ is the energy density for the point of transition. Moreover, 
$\Delta\epsilon$ is the discontinuity of the energy density at the transition and $c_{\rm QM}$ is the speed of sound of quark matter (QM). In our analysis, we will assume $c_{\rm QM}^2 = 1$ in order to achieve the stiffest possible EoS. The resulting predictions are dependent on the values assumes for $\Delta\epsilon$ and $p_t$, with the twin star solutions being generated for some combinations of these parameters. The studies performed in Refs.~\cite{Christian:2017jni,Christian:2020xwz} have demonstrated that the maximum mass of the second branch is determined by $\Delta\epsilon$, while  the maximum mass of the first branch is influenced by $p_t$.  Moreover, these authors have classified the twin star solutions in four categories, depending on the masses of the twin stars, which are defined by: (a) Category I: the maximum masses of both stars are larger than $\SI{2}{M_{\odot}}$; (b) Category II: only the first maximum is larger than $\SI{2}{M_{\odot}}$; (c) Category III: the first maximum is in the range  $1 \le M_{\mathrm{Had}}/\si{M_{\odot}} \le 2$, while the second one is larger than $\SI{2}{M_{\odot}}$; and (d) Category IV: the first maximum is smaller than  $\SI{1}{M_{\odot}}$, while the second one is larger than $\SI{2}{M_{\odot}}$. In Fig.~\ref{fig:exp-MxR-4categories} (left panel), we present the typical EoSs for these four categories, with the $(\Delta\epsilon, \, p_t)$ values being given in the parenthesis. For completeness, we also present, in the right panel, the resulting  predictions for the mass-radius relations derived by solving the TOV equation for a neutral compact star. The pure hadronic case (black thin curve) obtained from the Hebeler {\it et al.} stiffest EoS is presented for comparison. For the case where a phase transition is present, the maximum hadronic and hybrid masses are marked by filled circles. Moreover, the dotted lines indicate the unstable configurations. As it is clear from the figure, two separate branches, characteristic of twin stars, are present.

\begin{figure*}[!tb]
  \centering
  \includegraphics[width=.45\textwidth]{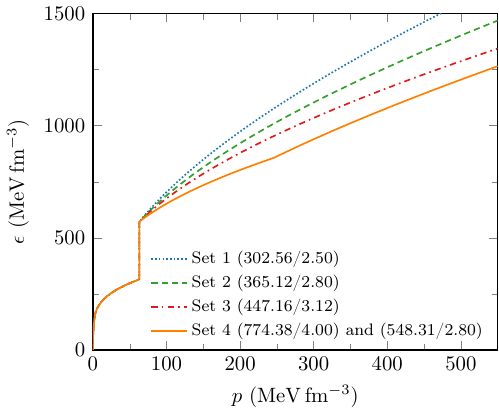}\hfill
  \includegraphics[width=.43\textwidth]{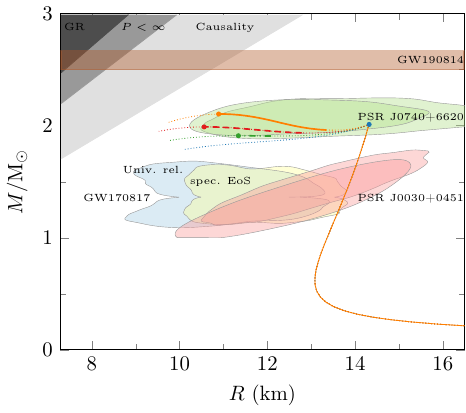}
  \caption{\label{fig:Alvarez-exp-MxR} {\it Left panel:} EoSs for the MP model considering  four sets of values for the $\kappa$ and $\Gamma$ parameters (in parenthesis) that characterize quark matter in the high density regime. As in Ref.~\cite{Alvarez-Castillo:2017qki}, for Set 4, we describe the high density regime in terms of two polytrope branches, depending on the nuclear density.  {\it Right panel:} Corresponding predictions for the mass-radius relation of a neutral compact star. The maximum hadronic and hybrid masses are marked by filled circles. The same constraints and limits of Fig.~\ref{fig:exp-MxR-4categories} are presented.}
\end{figure*}

In our analysis, we also  consider an  EoS for QM with a speed of sound dependent on the density. Inspired by  the study performed in Ref.~\cite{Alvarez-Castillo:2017qki}, we will assume that high-density region can be described by a polytrope EoS given by $P(n) = \kappa n^{\Gamma}$, where $n$ is the nuclear density and $\Gamma$  is the polytrope parameter. As in Ref.~\cite{Alvarez-Castillo:2017qki}, we will consider four different sets of parameters for $\kappa$ and $\Gamma$, with the maximum values of $\Gamma$ being determined by imposing that the speed of sound does not exceed the speed of light for the values of density reached in the center of the star. As in the CSS case, the hadronic phase will be described by the stiffest EoS of Hebeler {\it et al.}~\cite{hebeler2013}. For all sets of parameters for the QM EoS, we assume that the critical pressure is $p_{t} = \SI{63.177}{MeV\,fm^{-3}}$, that the discontinuity of the energy density at the transition is $\Delta \epsilon = \SI{253.89}{MeV\,fm^{-3}}$ \,{and that the transitional baryon number density is 2.03 $n_0$}, in agreement with the previous estimation of these values\footnote{This value of transitional pressure ensures hadronic stars with $\SI{2}{M_\odot}$ masses. Furthermore, the jump in energy density is related to the latent heat of the (assumed) first-order QCD transition being estimated from the $\Lambda_{\rm QCD}$ scale~(see, e. g. Refs.~\cite{kurkela2014,fraga2015})}. The EoSs for the MP model are presented in Fig. \ref{fig:Alvarez-exp-MxR} (left panel), while the corresponding predictions for the mass-radius relation are shown in the right panel, which were derived assuming a neutral compact star. One has that they share the same  hadronic branch, and that the maximum mass of the second one increases for larger values of $\Gamma$. In particular, one has that for Set 1, the second branch is unstable and twin star solutions are not obtained. On the other hand, a third family of stable stars is present for the other sets, in agreement with the results obtained in Ref.~\cite{Alvarez-Castillo:2017qki}.

\section{Results}
\label{sec:res}

In what follows, we will analyze the impact of the electric charge on the results derived in the previous section for the neutral twin stars predicted by the CSS and MP models. In general terms, one has for both models that the presence of an electric charge distribution implies  stellar configurations that have a larger mass and radius compared to its neutral counterparts. However, depending on the amount of the electric charge, the new configurations become unstable and twin stars solutions are not present. Another important aspect for the case of stable charged twin stars configurations, is that the gap mass between the two branches is sensitive to the value of $\alpha$, which establishes the relation between the electric charge distribution and the energy density ($\rho_e = \alpha\epsilon$).
Finally, our calculations indicate that both models are able to predict stable charged twin stars with masses larger than $\SI{2.5}{M_{\odot}}$ in contrast to charged hybrid stars with Gibbs transitions like in Ref. \cite{brillante2014}.

\begin{figure}[t]
  \centering
   \includegraphics[width=.35\textwidth]{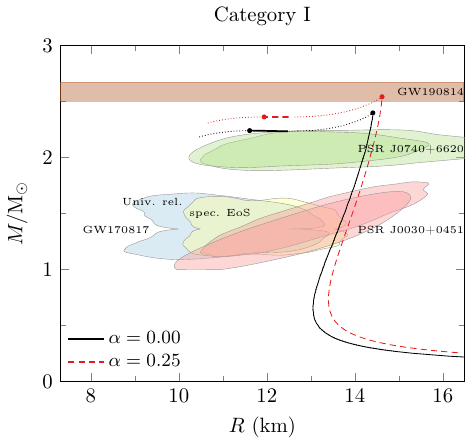}
  \includegraphics[width=.35\textwidth]{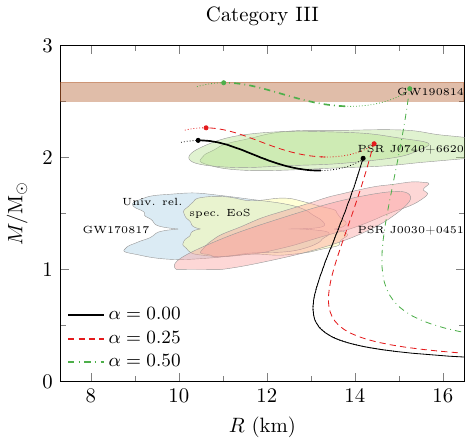}\hfill
  \includegraphics[width=.35\textwidth]{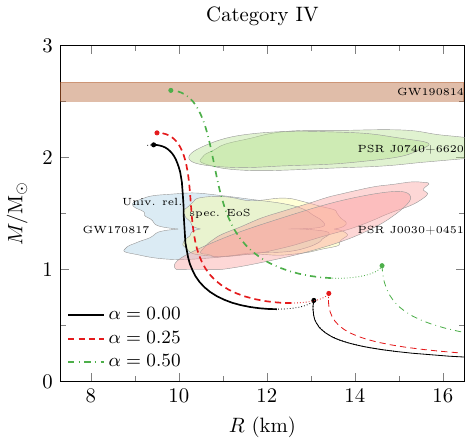}
  \caption{\label{fig:MxR-categoryIII-IV-alpha} Predictions of the CSS model for the  mass-radius profile of the charged twin stars considering the Categories I, III and IV. The results for the neutral counterparts are also presented for comparison.}
\end{figure}

Initially, in Fig.~\ref{fig:MxR-categoryIII-IV-alpha} we present our results for the distinct categories of the CSS model and different values of $\alpha$. The neutral solution is also presented for comparison. The filled circles represent the maximum mass configurations for the hadronic and hybrid cases. We will focus on the values of  $\alpha \ge 0.25$, where we expect a larger impact of the electric charge on the stellar configuration~\cite{lazzari2020}. As a consequence, we will not show the results for  twin  stars configurations classified as Category II, since we have verified that they do not present a hybrid branch for $\alpha \ge 0.2$. 
The results presented in Fig.~\ref{fig:MxR-categoryIII-IV-alpha} demonstrate that the presence of charge increases the masses of the twin stars in both branches of all categories considered. However, 
we have found that Categories I, II and III do not present twin stable configurations for large values of $\alpha$. In particular, for Category I,  the hybrid branch is unstable for  $\alpha > 0.25$, even though the hadronic mass reaches the hypothetical constraint of GW190814 in the charged case. Moreover, for $\alpha = 0.25$, the hybrid branch becomes smaller in comparison to the neutral counterpart. 
On the other hand, our results indicate that the hybrid branch becomes unstable for $\alpha > 0.8$ for Category III, while Category IV has stable twin star solutions for the largest value of $\alpha$ allowed (close to unity), \,{but that do not satisfy any of the previous mass-radius constraints}.

\,{A few comments are in order. Regarding the GW170817 mass-radius constraint, it is only satisfied for Category IV twin stars. The remaining constraints are in general satisfied in every other category for larger values of alpha (when mentioned) for stable charged twin stars. In particular, the NICER constraints for PSR J0030+0451 at 68\% confidence interval are satisfied by all of our $\SI{1.4}{M_\odot}$ charged stellar models which represent hadronic stars (except for Category IV, where $\SI{1.4}{M_\odot}$ stars are hybrid ones).}


In Table \ref{tab:categoryI}, we present our predictions for the maximum gravitational masses of the hadronic (Had) and hybrid (Hyb) configurations as well as the corresponding radii for the distinct categories of the CSS model. \,{We have also included the radius of $\SI{1.4}{M_\odot}$ stars for comparison with the NICER results discussed above.}
In agreement with the previous discussions, the  maximum masses and radii increase for larger values of $\alpha$. In particular, for Categories III and IV, the hybrid branches for $\alpha = 0.5$ clearly fulfill the hypothetical GW190814 mass constraint and is easily surpassed for larger values of $\alpha$ (not shown). However, the hadronic maximum mass of category IV barely surpasses $\SI{1}{M_\odot}$ even in the charged case. An interesting aspect is that the radius difference  of the twin star solutions for category IV becomes larger than 4.5 km for $\alpha \ge 0.5$.

\begin{table*}[t]
    \centering
    \begin{tabular}{c||c||c|c||c||c|c}
  \bf{Category} &     \bf{ $\alpha$} & \bf{ $M_{\mathrm{Had}}^{\mathrm{Max}} [\mathrm{M}_\odot]$} & \bf{ $R_{\mathrm{Had}} [\si{km}]$} & \bf{ $R_{1.4} [\si{km}]$} & \bf{$M_{\mathrm{Hyb}}^{\mathrm{Max}} [\mathrm{M}_\odot]$} & \bf{$R_{\mathrm{Hyb}} [\si{km}$]}\\
        \hline
        \hline
   \multirow{2}{*}{I} &     0.00 & 2.40 & 14.41 & 13.67 & 2.24 & 11.67\\
     &   0.25 & 2.54 & 14.62 & 13.80 & 2.36 & 11.94 \\ 
   \hline
        \hline
 &         0.00 & 1.99 & 14.19 &  13.67 & 2.15 & 10.44\\
III &        0.25 & 2.12 & 14.44 & 13.80 & 2.27 & 10.62 \\
 &        0.50 & 2.61 & 15.25 & 14.70 & 67 & 11.02\\
   \hline
        \hline
        
      &      0.00 & 0.73 & 13.07 & 10.12 & 2.11 & 9.43\\
      IV &   0.25 & 0.79 & 13.41 & 10.31 & 2.22 & 9.50 \\
        &  0.50 & 1.04 & 14.62 & 11.19 & 2.60 & 9.82\\
   \hline
        \hline
    \end{tabular}
    \caption{Maximum gravitational masses for the hadronic (Had) and hybrid (Hyb) configurations with corresponding radii for the distinct categories of the CSS model.}
    \label{tab:categoryI}
\end{table*}

In order to verify the robustness of our findings, we also have studied the stellar configurations assuming that the quark EoSs are characterized by  $c_{\mathrm{QM}}^2 = 0.33$ and 0.7. In both scenarios, we have found that the charged hybrid branch is completely unstable for Categories I and II for all values of $\alpha$. Category III has unstable hybrid configurations for $c_{\mathrm{QM}}^2 = 0.33$, but has a stable one for $c_{\mathrm{QM}}^2 = 0.7$. In the latter, only the hadronic configurations surpass the $\SI{2.6}{M_\odot}$ limit. For Category IV, where the phase transition pressure is small, we have found a stable hybrid branch for both speeds of sound. Finally,  $M \approx \SI{2.6}{M_\odot}$  is reached for $c_{\mathrm{QM}}^2 = 0.33 \, (0.7)$ when $\alpha \geq 0.8\, (0.6)$. 

\begin{figure}[t]
  \centering
  \includegraphics[width=.35\textwidth]{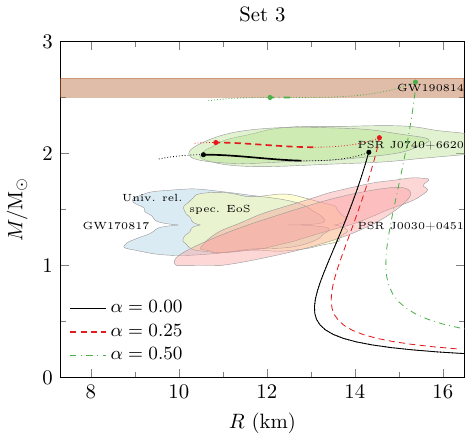}\hfill
  \includegraphics[width=.35\textwidth]{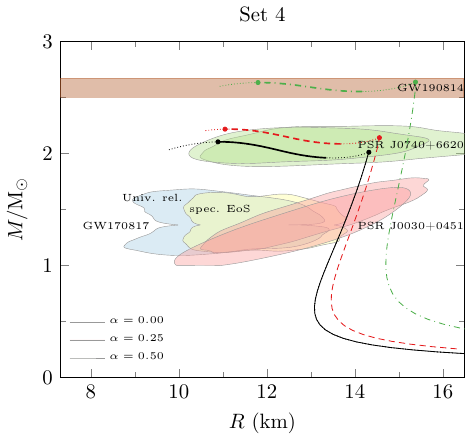}
  \caption{\label{fig:MxR-Set3-4-alpha} Predictions of the MP model for the mass-radius profile of charged twin stars considering sets 3 and 4. The results for the neutral counterparts are also presented for comparison. }
\end{figure}

\begin{table*}[t]
    \centering
    \begin{tabular}{c||c||c|c||c|c|c}
    {\bf{Set}}&      \bf{ $\alpha$} & \bf{ $M_{\mathrm{Had}}^{\mathrm{Max}} [\mathrm{M}_\odot]$} & \bf{ $R_{\mathrm{Had}} [\si{km}]$} & \bf{$M_{\mathrm{Hyb}}^{\mathrm{Max}} [\mathrm{M}_\odot]$} & \bf{$R_{\mathrm{Hyb}} [\si{km}$]} & \bf{$M_{\mathrm{Hyb}}^A~[\mathrm{M}_\odot]$}\\
        \hline
        \hline
 &        0.00 & 2.01 & 14.32 & 1.99 & 10.56 & 2.31 \\
3 &        0.25 & 2.14 & 14.56 & 2.10 & 10.84 & 2.41 \\
 &        0.50 & 2.64 & 15.38 & 2.50 & 12.07 & 2.77 \\\hline
\hline
      &        0.00 & 2.01 & 14.32 & 2.10 & 10.89 & 2.47 \\
4 &        0.25 & 2.14 & 14.56 & 2.22 & 11.05 & 2.57\\
 &        0.50 & 2.64 & 15.38 & 2.63 & 11.80 & 2.94\\\hline
\hline
    \end{tabular}
    \caption{Maximum gravitational masses for the hadronic (Had) and hybrid (Hyb) configurations with corresponding radii for distinct sets of the MP model. We also present the baryonic mass $M^A$ for the hybrid configurations.}
    \label{tab:Set3}
\end{table*}


Now we pass to estimate the impact of the electric charge distribution on the predictions of the MP model for the different sets. In the previous section, we have shown that stable twin star solutions are not obtained for neutral stars when Set 1 is considered. One has verified that such conclusion is also valid for the charged case. Moreover, we have obtained that for Set 2 the stable twin star configurations only occur for small values of the electric charge ($\alpha \le 0.1$). In contrast, such configurations are present in the Sets 3 and 4  for $\alpha \le $ 0.6 and 0.7, respectively. The corresponding predictions for the mass-radius relations of these sets are presented in Fig.~\ref{fig:MxR-Set3-4-alpha}. The results for the neutral counterparts are shown for comparison.
Our results indicate that for the Set 3, the GW190814 hypothetical mass constraint is satisfied by the hadronic branch for $\alpha = 0.5$, with the associated hybrid branch barely reaching the lower limit. For $\alpha = 0.6$, both branches satisfy the constraint. On the other hand, for larger values of $\alpha$,  all hybrid branches are unstable. In the case of Set 4, we have verified that the GW190814 hypothetical constraint is fulfilled for $\alpha = 0.5$, and stable charged twin star configurations, with larger masses and radii, are also present for $\alpha \le 0.7$. In contrast,  we have obtained that for $\alpha > 0.7$ there are no stable hybrid branches.

In Table~\ref{tab:Set3}, we present our predictions for the maximum gravitational masses of the hadronic (Had) and hybrid (Hyb) configurations as well as the corresponding radii for the Sets 3 and 4 of the MP model. \,{We have not included the radius of $\SI{1.4}{M_\odot}$ stars in Table~\ref{tab:Set3} because both Sets share the same hadronic configurations as the CSS model. So the $R_{1.4}$ are the same as the ones presented in Table~\ref{tab:categoryI} for categories I and III.} Similarly to what was observed for the CSS model, the electric charge also implies in larger maximum masses and radii for larger values of $\alpha$. An interesting aspect present in our predictions for Set 4 is that the impact of the electric charge on the hadronic and hybrid branches are distinct and dependent on the value of $\alpha$, with increasing of the hadronic maximum mass being faster than the hybrid one. One has that for $\alpha < 0.5$ the hybrid maximum mass is larger than the hadronic one, but for $\alpha \geq 0.5$ the hadronic maximum mass is already larger than the hybrid one. Finally, we have also calculated the baryon mass and compactness of charged twin stars considering the Set 4 and found that the charged stellar models also possess a negative binding energy, i.e. the baryon mass is larger than the gravitational one. Moreover, one has verified that the compactness of our charged twin star models are far below the `charged Buchdahl limit' presented in Ref.~\cite{andreasson2009} and, in fact, are even below 1/2 corresponding to Schwarzschild black holes. Both of these criteria serve to further verify the overall stability of our twin star configurations.

\section{Summary}\label{Conclusion}
The description of compact stars have been subject of intense attention in recent years, mainly motivated by the beginning of the multi-messenger era in astrophysics, which has provided new constraints on the bulk properties of matter inside these objects and high precision data for their masses and radii is becoming available. 
Considering this perspective, the nuclear community is currently challenged to improve the modelling of complex processes, as for example,  possible phase transitions in the  core matter and the coalescence of binary compact objects.
Our goal in this paper was to contribute to this effort by investigating, for the first time, the impact of an electric charge distribution on the stellar properties and stability of twin hybrid neutron stars. In particular, we have extended for the charged case the studies performed in Refs.~\cite{Alvarez-Castillo:2017qki,Christian:2020xwz}, which have determined limitations on the masses of stable twin stars configurations assuming distinct assumptions for the phase transition which we emulate in order to establish our findings within a well-known framework. In particular, our calculations indicate that in general, the presence of large values of electric charge imply twin star configurations with larger masses than the 
\,{maximum mass estimates made from gravitational-wave data~\cite{margalit2017,rezzolla2018}} and charged hybrid stars with Gibbs transitions \cite{brillante2014}. We have also verified that twin stars  with masses of $\approx \SI{2.6}{M_{\odot}}$ are stable in the lines of Ref.~\cite{pereira2018}, which allow us to interpret the unknown compact object measured in the GW190814 event as a charged NS instead of a very light BH. However, such interpretation is still a theme of debate and should only be considered an interesting coincidence with our generic findings. Finally, our results motivate us to extend our analysis on the formation of stable charged twin hybrid stars which, according to Ref.~\cite{Espino:2021adh}, are difficult to occur in the zero-charge limit.




\begin{acknowledgments}
The authors thank J. Schaffner-Bielich for useful discussions. This work was partially supported by INCT-FNA (Process No. 464898/2014-5). VPG and LL acknowledges support from CNPq, CAPES (Finance Code 001), and FAPERGS. JCJ acknowledges support from FAPESP (Processes No. 2020/07791-7 and No. 2018/24720-6).
\end{acknowledgments}\


\end{document}